\documentclass[aps,prl,floatfix,twocolumn,superscriptaddress,showpacs,amsmath,amssymb]{revtex4}
\usepackage{graphicx}
\usepackage{epsfig}

\begin{document}

\newcommand{\m}[1]{\mathcal{#1}} 
\newcommand{\eps}{\varepsilon}
\newcommand{\eqn}[1]{(\ref{#1})} 
\newcommand{\be}{\begin{equation}}
\newcommand{\ee}{\end{equation}} 
\newcommand{\bea}{\begin{eqnarray}}
\newcommand{\eea}{\end{eqnarray}} 
\newcommand{\ba}{\begin{eqnarray*}}
\newcommand{\ea}{\end{eqnarray*}}
\newcommand{\dagga}{{\phantom{\dagger}}}
\newcommand{\fract}[2]{\frac{\displaystyle #1}{\displaystyle #2}}

\title{Time-Dependent Mean Field Theory for Quench Dynamics in correlated electron systems} 
\author{Marco Schir\'o} 
\affiliation{International
  School for Advanced Studies (SISSA), and CRS Democritos, CNR-INFM,
  Via Beirut 2-4, I-34014 Trieste, Italy} 
\author{Michele Fabrizio} 
\affiliation{International School for
  Advanced Studies (SISSA), and CRS Democritos, CNR-INFM, Via Beirut
  2-4, I-34014 Trieste, Italy} 
\affiliation{The Abdus Salam
  International Centre for Theoretical Physics (ICTP), P.O.Box 586,
  I-34014 Trieste, Italy}

\date{\today} 

\begin{abstract}
A simple and very flexible variational approach to the out-of-equilibrium quantum dynamics 
in strongly correlated electron systems is introduced through a time-dependent Gutzwiller wavefunction. 
As an application, we study 
the simple case of a sudden change of the interaction in the fermionic Hubbard model and 
find at the mean field level an extremely rich behaviour. In particular, 
a dynamical transition between small and large quantum quench regimes is found to occur at 
half-filling, in accordance with the analysis of Eckstein {\sl et al.}, Phys. Rev. Lett. {\bf 103}, 
056403 (2009), obtained by dynamical mean field theory, that turns into a crossover at any finite doping.  
\end{abstract}
\maketitle

\textit{Introduction.}  
Triggered by the enormous advances in the physics of ultra-cold atomic gases,\cite{Bloch_rmp08} 
time dependent non equilibrium phenomena in strongly interacting quantum systems have recently 
become of greatest interest. The possibility of artificially engineering 
many-particle quantum states with tunable interactions and almost perfect isolation 
from the environment gives the chance of probing directly in the time domain the quantum dynamics 
following an external perturbation~\cite{Greiner_nature_02}.
While early experiments focus mainly on bosonic systems~\cite{Weiss_nature06} or 
fermionic condensates,\cite{Zwierlein_prl05} the recent experimental realization of 
a fermionic Mott insulator~\cite{Fermionic_Mott_08} opens the way to investigate
out-of-equilibrium phenomena in electron systems too.\cite{Demler_doublons_prl10} 
From a theoretical perspective, these experiments raise several intruiguing questions touching 
quantum dynamics at its roots. Indeed, when driven out of equilibrium, interacting quantum systems 
can display peculiar dynamical behaviours or even be trapped into metastable 
configurations~\cite{Rosch_prl08} that differ completely from their equilibrium counterpart. 
The simplest way one could imagine to induce a non trivial dynamics is through a so called 
\emph{quantum quench}. Here the system is firstly prepared in the ground-state of some given 
Hamiltonian $\m{H}_i$, and then suddenly let evolved under the action of a new hamiltonian $\m{H}_f$. 
Recently, quantum quenches have been the subject of a vast literature focusing on 
integrable systems,\cite{Rossini_prl09} one 
dimensional models,\cite{onedim} or systems close to 
a quantum criticality.\cite{DeGrandi_prb09} The interest on these class of non equilibrium problems 
relies both on the dynamics itself,\cite{Peter_prl09} as well as on the long-time properties where the 
issue of thermalization or its lack of is still highly 
debated.\cite{Olshanii_Rigol_nature,Biroli_Corinna_arXiv} For what concerns strongly correlated 
electrons in more than one dimension, the subject is still largely unexplored and progresses 
have been done only very recently. The single band fermionic Hubbard model is likely the simplest 
lattice model of correlated electrons emboding the competition between metallic and insulating 
behaviour driven by a local Hubbard repulsion $U$. Its Hamiltonian reads
\be\label{eqn:hubb}
\m{H}\left(t\right)= -\sum_{\sigma}\sum_{\langle i,j\rangle}\,t_{ij}\,
 c^{\dagger}_{i\sigma}\,c^\dagga_{j\sigma}+U\left(t\right)\sum_{i}\,
n_{i\uparrow}\,n_{i\downarrow}\,.
\ee
In two pioneering works,\cite{Kehrein_prl08,Werner_prl09} the response of a Fermi sea to a sudden 
switch-on of the Hubbard $U$ has been studied in infinite dimensions using respectively the 
flow-equation method and the Dynamical Mean Field Theory (DMFT). Results suggest the existence of 
two different regimes in the real-time dynamics depending on the final interaction strength $U_f$. 
At weak coupling,\cite{Kehrein_prl08} the systems is trapped at long-times into a quasi-stationary 
regime where correlations are more effective than in equilibrium. This \emph{pre-thermalization} 
phenomenon has been confirmed by DMFT results,\cite{Werner_prl09} which further indicate 
a true dynamical transition above a critical $U_{fc}$ towards another regime with pronounced oscillations 
in the dynamics of physical quantities. These intriguing results have been so far restricted 
to a quench starting from a non interacting system ($U_i=0)$ and, more importantly, 
limited to rather short accessible time scales and weak quenches, thus leaving open many important issues.

A simple and flexible approach, although less rigorous than e.g. DMFT, is thus extremely desirable 
and this is actually the aim of the present work. 
Specifically, here we propose a variational approach to the out-of-equilibrium 
dynamics of a correlated electron system based on a proper extension of the Gutzwiller wavefunction.  
We apply this technique to study the interaction quench in the Hubbard model, where we find a 
rich behaviour featuring a transition in the real-time dynamics at a critical quench 
line $U_{fc}\left(U_i\right)$, in accordance with Ref.~\cite{Werner_prl09} at $U_i=0$.  
Remarkably, a finite doping completely washes out this transition, leaving behind only a crossover 
from weak to strong coupling.

\textit{Variational Approach to Quantum Dynamics.}  Many basic concepts in the theory of 
strongly correlated systems, like e.g. the Brinkman-Rice scenario for the Mott transition,
have been originated from calculations based on a very simple and physically transparent variational approach 
introduced in the 60th's by Gutzwiller \cite{Brinkman_Rice}.
This approach has been so far applied only at equilibrium 
or at most in the linear response regime,\cite{Lorenzana} but it turns out to be so flexible to allow for 
full out-of-equilibrium calculations. 

For simplicity, we assume initially a many-body wavefunction $|\Psi_0\rangle$, which, for times $t>0$, 
is let evolve with a Hamiltonian $\mathcal{H}$ that includes sizable on-site 
interactions. In the spirit of the Gutzwiller approach, we make the following variational 
ansatz for the time-dependent wavefunction $|\Psi_{exact}(t)\rangle = \mathrm{e}^{-i\mathcal{H}t}\,|\Psi_0\rangle$
\be
|\Psi_{exact}(t)\rangle \simeq |\Psi(t)\rangle = 
\prod_i \mathrm{e}^{-i\mathcal{S}_i(t)}\, \m{P}_i(t) \,|\Phi(t)\rangle,\label{eqn:ansatz}
\ee
where $|\Phi(t)\rangle$ is a time-dependent Slater determinant. 
$\m{P}_i(t)$ is a hermitean operator that acts on the Hilbert space of site $i$ and controls  
the weights of the local electronic configurations. $\m{S}_i(t)$ is also hermitean and we assume it  
depends on some variables $\phi_{i\alpha}(t)$ such that 
\[
\fract{\partial}{\partial \phi_{i\alpha}}\,\mathrm{e}^{-i\m{S}_i} = 
-i\,\m{O}_{i\alpha}\,\mathrm{e}^{-i\m{S}_i},
\]
where $\m{O}_{i\alpha}$ is any local hermitean operator. 
Since \eqn{eqn:ansatz} is just a variational ansatz, it does not solve the full Schr{\oe}dinger equation. 
Our proposal is to determine the variational parameters by requiring: ({\sl i}) that the Heisenberg 
equations of motion of the local operators $\m{O}_{i\alpha}$ are satisfied when averaging 
over \eqn{eqn:ansatz}; ({\sl ii}) that the 
average energy $E = \langle \Psi(t)|\,\m{H}\,|\Psi(t)\rangle$ is, as it should be, 
conserved during the evolution.  Since, by definition, 
\[
\fract{\partial}{\partial \phi_{i\alpha}}\, \mathrm{e}^{i\m{S}_i}\,\m{H}\,\mathrm{e}^{-i\m{S}_i}\,
= i\,\mathrm{e}^{i\m{S}_i}\,\left[\m{O}_{i\alpha},\m{H}\right]\,\mathrm{e}^{-i\m{S}_i},
\]
it follows that 
\bea
\fract{\partial O_{i\alpha}}{\partial t} &=& -i 
\langle \Psi_{exact}(t)|\,\left[\m{O}_{i\alpha},\m{H}\right]\,|\Psi_{exact}(t)\rangle \nonumber\\
&& \equiv -i \langle \Psi(t)|\left[\m{O}_{i\alpha},\m{H}\right]|\Psi(t)\rangle =
-\fract{\partial E}{\partial \phi_{i\alpha}},\label{eqn:ansatz-i}
\eea
where the equivalence is our variational assumption. Within the Gutzwiller approximation,\cite{dimerboss} 
which is exact in the limit of infinite coordination lattices, 
$E = \langle \Phi(t)|\,\m{H}_*(t)\,|\Phi(t)\rangle$, where $\m{H}_*(t)$ is  
a non-interacting Hamiltonian that depends on all time-dependent variational 
parameters defining $\m{P}_i$ and $\m{S}_i$. In general, these parameters can be expressed in terms 
of $\phi_{i\alpha}$ and $O_{i\alpha}$.   
If we impose that $|\Phi(t)\rangle$ is the solution of the Schr{\oe}dinger equation, namely that
$-i\partial_t\,|\Phi(t)\rangle = \m{H}_*(t)\,|\Phi(t)\rangle$
and furthermore that  
\be
\fract{\partial \phi_{i\alpha}}{\partial t} = \fract{\partial E}{\partial O_{i\alpha}},\label{eqn:ansatz-ii}
\ee
conservation of energy follows automatically. Therefore, $\phi_{i\alpha}$ and $\m{O}_{i\alpha}$  
act like conjugate variables and the energy functional $E$ as their effective Hamiltonian.

As a simple application of the above variational scheme, we assume $\mathcal{H}$ to be the 
Hubbard model \eqn{eqn:hubb} at half-filling with $U(t\leq 0)=U_i\geq 0$ and $U(t>0)=U_f>U_i$,    
and furthermore we limit our analysis to homogeneous paramagnetic wavefunctions. In the limit of infinite 
coordination lattices, one can compute exactly average values on the variational wavefunction 
provided the following conditions are imposed~\cite{dimerboss,Gebhard}
\[
\langle \Phi(t)\vert\m{P}^2_i(t)\vert\Phi(t) \rangle = 1\,,\, \langle
\Phi(t)\vert\m{P}^2_i(t)\,c^{\dagger}_{i\sigma}c^\dagga_{i\sigma}\vert\Phi(t)
\rangle = \frac{1}{2}\,. 
\]
We assume $\m{P}_i(t) = \sum_{n=0}^2\,\lambda_{i,n}(t)\,\m{P}_{i,n}$, where 
$\m{P}_{i,n}$ is the projector at site $i$ onto configurations with $n=0,\dots,2$ electrons and 
$\m{S}_i(t) = \sum_{n=0}^2\,\phi_{i,n}(t)\,\m{P}_{i,n}$, which implies that $\phi_{i,n}(t)$ plays the role 
of the conjugate variable of the occupation probability $P_{i,n} = \langle \Psi(t)\vert \m{P}_{i,n} 
\vert \Psi(t)\rangle$. From the constraints above it follows that $P_{i,0}=P_{i,2}$ and $P_{i,1}=1-2P_{i,2}$. 
We define $P_{i,2}=(1-\cos\theta_i)/4$ and set $\phi_{i,0}=\phi_{i,2}=\phi_i$ while $\phi_{i,1}=0$. 
Using $\theta_i$ and $\phi_i$ as variational 
parameters, one finds the average energy~\cite{dimerboss}
\bea
E &=& \fract{U_f}{4}\sum_{i}\left(1-\cos\theta_i(t)\right) \label{eqn:E-var} \\
&& - \sum_{ij}w_{ij}(t)\,\sin\theta_i(t)\,\cos\phi_i(t)\,
\sin\theta_j(t)\,\cos\phi_j(t)\,,\nonumber
\eea
where $w_{ij}(t)=t_{ij}\,\sum_\sigma\,\langle
\Phi(t)\vert\,c^{\dagger}_{i\sigma}c^\dagga_{j\sigma}+H.c.\,\vert\Phi(t) \rangle$. One recognizes 
in \eqn{eqn:E-var} the mean field energy of an Ising model in a transverse field
\be
\mathcal{H}_{I} = \fract{U_f}{4}\sum_{i}\,\left(1-\sigma^z_i\right) - 
\sum_{ij}\,w_{ij}(t)\,\sigma^x_i\,\sigma^x_j,\label{eqn:Ising}
\ee
where $\langle \sigma^z_i\rangle = \cos\theta_i$ and $\langle \sigma^x_i \rangle = \sin\theta_i\,\cos\phi_i$. 
This connection can be established rigorously at the variational level,\cite{nostro_unpublished} and agrees 
with the $Z_2$-slave-spin theory recently introduced.~\cite{Z2-1,Z2-2}
Therefore, it is 
not surprising that the equations of motion that we obtain through Eqs.~\eqn{eqn:ansatz-i} and 
\eqn{eqn:ansatz-ii} are just those of the Ising model $\langle \partial_t\,\sigma^a_i\rangle 
= -i\langle \left[\sigma^a_i,\mathcal{H}_I\right]\rangle$ within mean field. 
Therefore, under the above assumption of 
homogeneous and paramagnetic wavefunctions, a quantum quench in the half-filled Hubbard model is 
equivalent, within the Gutzwiller variational scheme, to a quench in a Ising model in the presence of 
a transverse field. In particular, if $|\Phi(t)\rangle$ is taken to be the half-filled Fermi sea, then 
$w_{ij}(t)=w$ and \eqn{eqn:Ising} is the conventional ferromagnetic Ising model with constant and uniform 
exchange $w$ and transverse field $U_f/4$. Quantum quenches of the transverse field have been recently 
investigated in one-dimension \cite{Rossini_prl09}
and on a fully connected lattice \cite{Sengupta}.
In the following we assume that the system is prepared in the metallic 
variational wavefunction that optimizes \eqn{eqn:hubb} with $U=U_i< U_c$, 
where $U_c$ is the variational estimate of the Mott transition.
\begin{figure}[t]
\begin{center}
\epsfig{figure=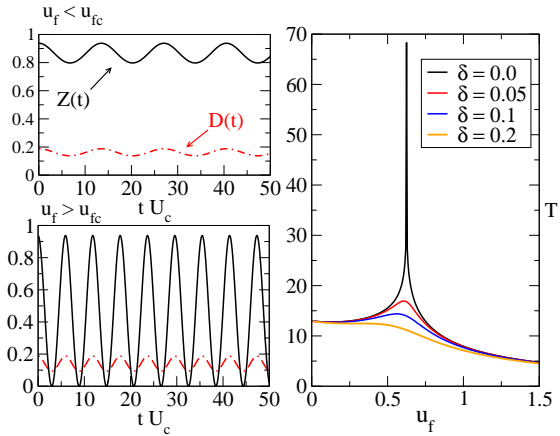,scale=0.31}
\caption{Left Panels: Gutzwiller mean field dynamics at half-filling 
for quasiparticle weight $Z(t)$ (black line) and double occupation $D(t)$ 
(dashed red line) for quantum quenches from $u_i=0.25$ to $u_f=0.35$ (top panel) and $u_f=1.25$ (bottom panel). Right Panel: Period of oscillations at 
half filling and for a finite doping $\delta\neq0$. 
Note the logarithmic singularity at $u_{fc}$ for $\delta=0$ (see main text) 
which is cut-off by finite doping.}
\label{fig:fig1}
\end{center}
\end{figure}
This corresponds to initial values $\phi_i(0)=0$ 
and $\cos\theta_i(0) = U_i/U_c\equiv u_i$ for the 
coupled equations: $2\dot{\phi} = U_c\,\cos\theta\cos^2\phi - U_f$ and 
$2\dot{\theta} = U_c\,\sin\theta\sin\phi\cos\phi$.
We note that, apart from the trivial case in which $U_f=U_i$, 
these equations admit a non-trivial 
stationary solution $\theta=0$ and $\cos^2\phi= U_f/U_c=u_f$, 
which is compatible with the initial conditions 
only when $u_f=u_{fc}=(1+u_i)/2$. 
It turns out that $u_{fc}$ identifies 
a dynamical critical point that separates two different regimes 
similarly to a simple pendulum. 
When $u_f<u_{fc}$, $2\phi(t)$ oscillates around the origin, while, 
for $u_f>u_{fc}$, it performs a cyclic motion around the whole circle.

In order to characterize the different regimes, 
we focus on three physical quantities, the double occupancy 
$D(t) = (1-\cos\theta(t))/4$, 
the quasiparticle residue $Z(t)= \sin^2\theta(t)\cos^2\phi(t)$ 
and their period of oscillation, $\m{T}$.    
\begin{figure}[t]
\begin{center}
\epsfig{figure=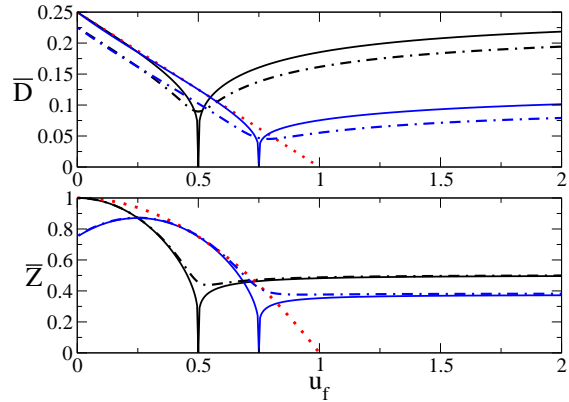,scale=0.3}
\caption{Average double occupation $\bar{D}$ (top) and quasiparticle 
weight $\bar{Z}$ (bottom) as a function of $u_f$ at fixed $u_i=0.0,0.5$. 
We show results at half filling (full lines) that display a sharp transition 
at $u_{fc}$, as well as at finite doping (dashed lines) where only a 
crossover remains. We also plot the zero temperature equilibrium 
results for $\delta=0$ (red points).}
\label{fig:fig2}
\end{center}
\end{figure}
While detailed calculations will be presented 
elsewhere~\cite{nostro_unpublished}, 
in the rest of the paper we just sketch the results of the mean field dynamics. 
Let us start from the weak coupling side $u_i<u_f< u_{fc}$, 
see top panel in Fig.~\ref{fig:fig1}, 
where both $D(t)$ and $Z(t)$ display small oscillations. 
Their amplitude and period \emph{increase} 
with the strength of the quench $\delta u = u_f-u_i$, 
the latter reading $\m{T}=\frac{4\sqrt{2}\,\m{K}\left(k\right)}{\sqrt{Z(0)}}$, 
where $\m{K}\left(k\right)$ is the 
complete elliptic integral of the first kind with 
argument $k^2=4u_f\delta u/Z(0)$. 
For $\delta u\to 0$ we find a linear increase 
$\m{T}\simeq\m{T}_0\left(1+u_i\,\delta u/Z(0)\right)$ 
with $\m{T}_0=4\pi/\sqrt{Z(0)}$.

Conversely, when quenching above the critical value, $u_f>u_{fc}$, 
a novel strong-coupling 
dynamical behaviour emerges. Here oscillations become faster, 
their period $\m{T}=4\m{K}\left(1/k\right)/\sqrt{u_f\delta u}$ 
now decreases as a function of $\delta u$. 
In particular, for $u_f\gg u_i$, we get $\m{T}\simeq\,\frac{2\pi}{u_f}$ 
smoothly matching the atomic 
limit result. The oscillation amplitude of $D(t)$ decreases with $u_f$, 
which results into a frozen dynamics in the infinite quench 
limit,\cite{Rosch_prl08} 
while quasiparticle weight $Z(t)$ still shows large oscillations 
even for $u_f\rightarrow\infty$, 
mainly reflecting the unbounded dynamics of the phase $\phi(t)$. 

Remarkably, the weak and the strong coupling regimes are 
separated by a critical quench 
line $u_{fc}$ at which mean-field dynamics exhibits 
\emph{exponential relaxation}. 
Indeed, upon approaching this line from both sides, 
the period $\m{T}$ diverges logarithmically, 
$\m{T}\simeq \frac{4}{\sqrt{Z(0)}}\,\mbox{log}\left(\frac{1}{\vert u_f-u_f^c\vert}\right)$. 
Right at criticality, $u_f=u_{fc}$, the mean field dynamics can be integrated 
exactly. 
The result gives 
$D(t)=D(0)\left(1-\mbox{tgh}^2\left(t/\tau_{\star}\right)\right)$;  
the double occupation relaxes exponentially to $\bar{D}=0$ pushing 
also $\bar{Z}\to 0$, 
with a characteristic time scale $\tau_{\star} = 2/\sqrt{Z(0)}$ 
that increases upon approaching the Mott Insulator. 

We now turn to discuss long-time average properties 
of the Gutzwiller mean-field dynamics that 
we define as 
$\bar{O}=\mbox{lim}_{t\rightarrow\infty}\frac{1}{t}\int_0^t\,dt'\,O(t')$. 
The analytical expressions~\cite{nostro_unpublished} 
of $\bar{D}$ and $\bar{Z}$ are shown in Fig.~\ref{fig:fig2}. 
At weak coupling, we find 
$\bar{D} = D(0)\left[1+\frac{u_f^c}{u_f}\left(\frac{E(k)-\m{K}(k)}{\m{K}(k)}
\right)\right]$, 
where $E\left(k\right)$ is the complete elliptic integral of 
the second kind and $k^2$ is the same 
as before. In addition, due to energy conservation, 
the knowledge of $\bar{D}$ 
completely fixes the average quasiparticle 
weight $\bar{Z}=Z(0)+8u_f\left(\bar{D}-D(0)\right)$.

It is interesting to consider first the small 
quench limit $\delta u\rightarrow0$. 
We find that, given $D(0) = (1-u_i)/4$ the initial equilibrium value,  
$\bar{D} \simeq D(0) -\delta u/4 = (1-u_f)/4$, 
namely tends to the equilibrium value corresponding to the 
final interaction.  Hence the prethermalization result~\cite{Kehrein_prl08} 
for the quasiparticle weight $\bar{Z}$ immediately follows. 
Indeed, quenching from 
a non-interacting Fermi sea, $u_i=0$ hence $Z(0)=1$, 
we find that the non-equilibrium $\bar{Z}$ is reduced twice more 
than its equilibrium value at $U_f$. 

For large quenches, the average double occupancy $\bar{D}$ \emph{increases} 
as a function of the final interaction $u_f$, 
$\bar{D} = D(0) -\frac{1}{2}(u_f-u_i)\left(1-E\left(1/k\right)/\m{K}\left(1/k\right)\right)$, 
eventually approaches its initial value $D(0)$ for $u_f\to\infty$. 
A similar behaviour is found for $\bar{Z}$ which is 
however further reduced by a factor $1/2$ 
with respect to the initial value $Z(0)$ due to the freely 
oscillating behaviour 
of the phase. We find therefore that, 
for large quantum quenches, the dynamics retains memory 
of the initial condition and thermalization is prevented by 
a dynamical blocking.

Finally, for quenches close to $u_{fc}$, both $\bar{D}$ and $\bar{Z}$ 
are very small, vanishing 
as $1/\log\vert u_f-u_{fc}\vert$ on approaching the critical point. 
Namely, $u_{fc}$ not only signals a transition in the dynamics but also 
identifies the critical interaction at which the quenched system shows 
genuine Mott insulating behaviour.

Away from half-filling, the dynamical equations become more 
cumbersome \cite{nostro_unpublished}. 
However, key features can be easily derived even without resorting to 
a numerical integration. In particular, we find that any finite doping 
turns the half-filled dynamical critical point into a crossover. 
For instance, the logarithmic singularity of the oscillatory period $\m{T}$ 
is cut-off by any finite doping, as shown in figure \ref{fig:fig1}.   
As a consequence, the singular behavior of the average values 
across the half-filling transition is smoothed into a crossover 
at finite doping, see figure \ref{fig:fig2}.

\textit{Discussion.} It is worth discussing the above results in light of those recently obtained by 
DMFT.\cite{Werner_prl09} Remarkably, our variational ansatz~(\ref{eqn:ansatz}) seems to catch many 
non trivial effects observed in DMFT. In particular the existence of two different 
regimes separated by a real dynamical transition at $u_{fc}$, already suggested in~\cite{Werner_prl09}, 
clearly emerges from our mean field theory. 

We note however that the suppression of quantum fluctuations, 
which is at the ground of our results, give rise to an oversimplified periodical dynamics that 
lacks relaxation.
In this respect DMFT, which can treat 
exactly all local quantum fluctuations, works much better and turns these oscillations 
into a true relaxation. We guess that a similar result could be obtained from our variational treatment, 
for example, by allowing fluctuations in the Fermi sea and treating the Quantum Ising Model beyond 
the simplest mean field level~\cite{Z2-1}. 
Finally we notice that the Ising analogy provides a simple 
interpretation of the dynamical transition, at least from a local DMFT-like point of view. Indeed, 
if we assume that the role of the neglected quantum fluctuations is to provide  
dissippation for the two-level system described by the local Ising variable, the two dynamical 
regimes found by DMFT resemble those in the phase diagram of the spin-boson 
model.\cite{spin-boson}   
%

\textit{Conclusion.} We introduced a variational approach to the out-of-equilibrium dynamics 
in strongly correlated electron systems. Using a time dependent Gutzwiller ansatz we address 
the problem of an interaction quench in the Hubbard model 
finding a dynamical transition at half-filling. Our results provide a simple and 
intuitive mean field theory for the quench dynamics in interacting Fermi systems.

\textit{Acknowledgments.} We acknowledge useful discussions with M. Capone, E. Tosatti, A. Millis, 
and C. Castellani. This work has been supported by Italian Ministry of University and Research, 
through a PRIN-COFIN award.  

\bibliographystyle{apsrev}

\begin{thebibliography}{28}
\expandafter\ifx\csname natexlab\endcsname\relax\def\natexlab#1{#1}\fi
\expandafter\ifx\csname bibnamefont\endcsname\relax
  \def\bibnamefont#1{#1}\fi
\expandafter\ifx\csname bibfnamefont\endcsname\relax
  \def\bibfnamefont#1{#1}\fi
\expandafter\ifx\csname citenamefont\endcsname\relax
  \def\citenamefont#1{#1}\fi
\expandafter\ifx\csname url\endcsname\relax
  \def\url#1{\texttt{#1}}\fi
\expandafter\ifx\csname urlprefix\endcsname\relax\def\urlprefix{URL }\fi
\providecommand{\bibinfo}[2]{#2}
\providecommand{\eprint}[2][]{\url{#2}}

\bibitem[{\citenamefont{Bloch et~al.}(2008)\citenamefont{Bloch, Dalibard, and
  Zwerger}}]{Bloch_rmp08}
\bibinfo{author}{\bibfnamefont{I.}~\bibnamefont{Bloch}},
  \bibinfo{author}{\bibfnamefont{J.}~\bibnamefont{Dalibard}}, \bibnamefont{and}
  \bibinfo{author}{\bibfnamefont{W.}~\bibnamefont{Zwerger}},
  \bibinfo{journal}{Rev. Mod. Phys.} \textbf{\bibinfo{volume}{80}},
  \bibinfo{pages}{885} (\bibinfo{year}{2008}).

\bibitem[{\citenamefont{Greiner et~al.}(2002)\citenamefont{Greiner, Mandel,
  Esslinger, Hansch, and Bloch}}]{Greiner_nature_02}
\bibinfo{author}{\bibfnamefont{M.}~\bibnamefont{Greiner}},
  \bibinfo{author}{\bibfnamefont{O.}~\bibnamefont{Mandel}},
  \bibinfo{author}{\bibfnamefont{T.}~\bibnamefont{Esslinger}},
  \bibinfo{author}{\bibfnamefont{T.}~\bibnamefont{Hansch}}, \bibnamefont{and}
  \bibinfo{author}{\bibfnamefont{I.}~\bibnamefont{Bloch}},
  \bibinfo{journal}{Nature} \textbf{\bibinfo{volume}{415}}, \bibinfo{pages}{39}
  (\bibinfo{year}{2002}).

\bibitem[{\citenamefont{Kinoshita et~al.}(2006)\citenamefont{Kinoshita, Wenger,
  and Weiss}}]{Weiss_nature06}
\bibinfo{author}{\bibfnamefont{T.}~\bibnamefont{Kinoshita}},
  \bibinfo{author}{\bibfnamefont{T.}~\bibnamefont{Wenger}}, \bibnamefont{and}
  \bibinfo{author}{\bibfnamefont{D.~S.} \bibnamefont{Weiss}},
  \bibinfo{journal}{Nature} \textbf{\bibinfo{volume}{440}},
  \bibinfo{pages}{900} (\bibinfo{year}{2006}).

\bibitem[{\citenamefont{Zwierlein et~al.}(2005)\citenamefont{Zwierlein,
  Schunck, Stan, Raupach, and Ketterle}}]{Zwierlein_prl05}
\bibinfo{author}{\bibfnamefont{M.~W.} \bibnamefont{Zwierlein}},
  \bibinfo{author}{\bibfnamefont{C.~H.} \bibnamefont{Schunck}},
  \bibinfo{author}{\bibfnamefont{C.~A.} \bibnamefont{Stan}},
  \bibinfo{author}{\bibfnamefont{S.~M.~F.} \bibnamefont{Raupach}},
  \bibnamefont{and} \bibinfo{author}{\bibfnamefont{W.}~\bibnamefont{Ketterle}},
  \bibinfo{journal}{Phys. Rev. Lett.} \textbf{\bibinfo{volume}{94}},
  \bibinfo{pages}{180401} (\bibinfo{year}{2005}).

\bibitem[{\citenamefont{Jordens et~al.}(2008)\citenamefont{Jordens, Strohmaier,
  Gunter, Moritz, and Esslinger}}]{Fermionic_Mott_08}
\bibinfo{author}{\bibfnamefont{R.}~\bibnamefont{Jordens}},
  \bibinfo{author}{\bibfnamefont{N.}~\bibnamefont{Strohmaier}},
  \bibinfo{author}{\bibfnamefont{K.}~\bibnamefont{Gunter}},
  \bibinfo{author}{\bibfnamefont{H.}~\bibnamefont{Moritz}}, \bibnamefont{and}
  \bibinfo{author}{\bibfnamefont{T.}~\bibnamefont{Esslinger}},
  \bibinfo{journal}{Nature} \textbf{\bibinfo{volume}{451}}
  (\bibinfo{year}{2008}).

\bibitem[{\citenamefont{Strohmaier~\textit{et
  al.}}(2010)}]{Demler_doublons_prl10}
\bibinfo{author}{\bibfnamefont{N.}~\bibnamefont{Strohmaier~\textit{et al.}}},
  \bibinfo{journal}{Phys. Rev. Lett.} \textbf{\bibinfo{volume}{104}},
  \bibinfo{pages}{080401} (\bibinfo{year}{2010}).

\bibitem[{\citenamefont{Rosch et~al.}(2008)\citenamefont{Rosch, Rasch, Binz,
  and Vojta}}]{Rosch_prl08}
\bibinfo{author}{\bibfnamefont{A.}~\bibnamefont{Rosch}},
  \bibinfo{author}{\bibfnamefont{D.}~\bibnamefont{Rasch}},
  \bibinfo{author}{\bibfnamefont{B.}~\bibnamefont{Binz}}, \bibnamefont{and}
  \bibinfo{author}{\bibfnamefont{M.}~\bibnamefont{Vojta}},
  \bibinfo{journal}{Phys. Rev. Lett.} \textbf{\bibinfo{volume}{101}},
  \bibinfo{pages}{265301} (\bibinfo{year}{2008}).

  
\bibitem[{\citenamefont{Rossini et~al.}(2009)\citenamefont{Rossini, Silva,
  Mussardo, and Santoro}}]{Rossini_prl09}
  \bibinfo{author}{\bibfnamefont{M.}~\bibnamefont{Eckstein}} \bibnamefont{and}
  \bibinfo{author}{\bibfnamefont{M.}~\bibnamefont{Kollar}},
  \bibinfo{journal}{Phys. Rev. Lett.} \textbf{\bibinfo{volume}{100}},
  \bibinfo{pages}{120404} (\bibinfo{year}{2008});
\bibinfo{author}{\bibfnamefont{D.}~\bibnamefont{Rossini}},
  \bibinfo{author}{\bibfnamefont{A.}~\bibnamefont{Silva}},
  \bibinfo{author}{\bibfnamefont{G.}~\bibnamefont{Mussardo}}, \bibnamefont{and}
  \bibinfo{author}{\bibfnamefont{G.~E.} \bibnamefont{Santoro}},
  \bibinfo{journal}{Phys. Rev. Lett.} \textbf{\bibinfo{volume}{102}},
  \bibinfo{pages}{127204} (\bibinfo{year}{2009}).
  

\bibitem[{\citenamefont{Cazalilla}(2006)}]{onedim}
\bibinfo{author}{\bibfnamefont{M.~A.} \bibnamefont{Cazalilla}},
  \bibinfo{journal}{Phys. Rev. Lett.} \textbf{\bibinfo{volume}{97}},
  \bibinfo{pages}{156403} (\bibinfo{year}{2006});
\bibinfo{author}{\bibfnamefont{C.}~\bibnamefont{Kollath}},
  \bibinfo{author}{\bibfnamefont{A.~M.} \bibnamefont{L\"{a}uchli}},
  \bibnamefont{and} \bibinfo{author}{\bibfnamefont{E.}~\bibnamefont{Altman}},
  \bibinfo{journal}{Phys. Rev. Lett.} \textbf{\bibinfo{volume}{98}},
  \bibinfo{pages}{180601} (\bibinfo{year}{2007});
\bibinfo{author}{\bibfnamefont{S.~R.} \bibnamefont{Manmana}},
  \bibinfo{author}{\bibfnamefont{S.}~\bibnamefont{Wessel}},
  \bibinfo{author}{\bibfnamefont{R.~M.} \bibnamefont{Noack}}, \bibnamefont{and}
  \bibinfo{author}{\bibfnamefont{A.}~\bibnamefont{Muramatsu}},
  \bibinfo{journal}{\emph{ibid}}, \bibinfo{pages}{210405}.


%

\bibitem[{\citenamefont{De~Grandi et~al.}(2010)\citenamefont{De~Grandi,
  Gritsev, and Polkovnikov}}]{DeGrandi_prb09}
\bibinfo{author}{\bibfnamefont{C.}~\bibnamefont{De~Grandi}},
  \bibinfo{author}{\bibfnamefont{V.}~\bibnamefont{Gritsev}}, \bibnamefont{and}
  \bibinfo{author}{\bibfnamefont{A.}~\bibnamefont{Polkovnikov}},
  \bibinfo{journal}{Phys. Rev. B} \textbf{\bibinfo{volume}{81}},
  \bibinfo{pages}{012303} (\bibinfo{year}{2010}).

\bibitem[{\citenamefont{Barmettler et~al.}(2009)\citenamefont{Barmettler, Punk,
  Gritsev, Demler, and Altman}}]{Peter_prl09}
\bibinfo{author}{\bibfnamefont{P.}~\bibnamefont{Barmettler}},
  \bibinfo{author}{\bibfnamefont{M.}~\bibnamefont{Punk}},
  \bibinfo{author}{\bibfnamefont{V.}~\bibnamefont{Gritsev}},
  \bibinfo{author}{\bibfnamefont{E.}~\bibnamefont{Demler}}, \bibnamefont{and}
  \bibinfo{author}{\bibfnamefont{E.}~\bibnamefont{Altman}},
  \bibinfo{journal}{Phys. Rev. Lett.} \textbf{\bibinfo{volume}{102}},
  \bibinfo{pages}{130603} (\bibinfo{year}{2009}).

\bibitem[{\citenamefont{Rigol et~al.}(2008)\citenamefont{Rigol, Dunjko, and
  Olshanii}}]{Olshanii_Rigol_nature}
\bibinfo{author}{\bibfnamefont{M.}~\bibnamefont{Rigol}},
  \bibinfo{author}{\bibfnamefont{V.}~\bibnamefont{Dunjko}}, \bibnamefont{and}
  \bibinfo{author}{\bibfnamefont{M.}~\bibnamefont{Olshanii}},
  \bibinfo{journal}{Nature} \textbf{\bibinfo{volume}{452}}
  (\bibinfo{year}{2008}).

\bibitem[{\citenamefont{Biroli et~al.}(2009)\citenamefont{Biroli, Kollath, and
  Lauchli}}]{Biroli_Corinna_arXiv}
\bibinfo{author}{\bibfnamefont{G.}~\bibnamefont{Biroli}},
  \bibinfo{author}{\bibfnamefont{C.}~\bibnamefont{Kollath}}, \bibnamefont{and}
  \bibinfo{author}{\bibfnamefont{A.~M.} \bibnamefont{Lauchli}},
  \bibinfo{journal}{arXiv:0907.3731v1}  (\bibinfo{year}{2009}).

\bibitem[{\citenamefont{Moeckel and Kehrein}(2008)}]{Kehrein_prl08}
\bibinfo{author}{\bibfnamefont{M.}~\bibnamefont{Moeckel}} \bibnamefont{and}
  \bibinfo{author}{\bibfnamefont{S.}~\bibnamefont{Kehrein}},
  \bibinfo{journal}{Phys. Rev. Lett.} \textbf{\bibinfo{volume}{100}},
  \bibinfo{pages}{175702} (\bibinfo{year}{2008}).

\bibitem[{\citenamefont{Eckstein et~al.}(2009)\citenamefont{Eckstein, Kollar,
  and Werner}}]{Werner_prl09}
\bibinfo{author}{\bibfnamefont{M.}~\bibnamefont{Eckstein}},
  \bibinfo{author}{\bibfnamefont{M.}~\bibnamefont{Kollar}}, \bibnamefont{and}
  \bibinfo{author}{\bibfnamefont{P.}~\bibnamefont{Werner}},
  \bibinfo{journal}{Phys. Rev. Lett.} \textbf{\bibinfo{volume}{103}},
  \bibinfo{pages}{056403} (\bibinfo{year}{2009}).
  

\bibitem[{\citenamefont{Brinkman and Rice}(1970)}]{Brinkman_Rice}
\bibinfo{author}{\bibfnamefont{W.~F.} \bibnamefont{Brinkman}} \bibnamefont{and}
  \bibinfo{author}{\bibfnamefont{T.~M.} \bibnamefont{Rice}},
  \bibinfo{journal}{Phys. Rev. B} \textbf{\bibinfo{volume}{2}},
  \bibinfo{pages}{4302} (\bibinfo{year}{1970});
\bibinfo{author}{\bibfnamefont{M.~C.} \bibnamefont{Gutzwiller}},
  \bibinfo{journal}{Phys. Rev.} \textbf{\bibinfo{volume}{137}},
  \bibinfo{pages}{A1726} (\bibinfo{year}{1965}).

\bibitem[{\citenamefont{Seibold and Lorenzana}(2001)}]{Lorenzana}
\bibinfo{author}{\bibfnamefont{G.}~\bibnamefont{Seibold}} \bibnamefont{and}
  \bibinfo{author}{\bibfnamefont{J.}~\bibnamefont{Lorenzana}},
  \bibinfo{journal}{Phys. Rev. Lett.} \textbf{\bibinfo{volume}{86}},
  \bibinfo{pages}{2605} (\bibinfo{year}{2001}).

\bibitem[{\citenamefont{Fabrizio}(2007)}]{dimerboss}
\bibinfo{author}{\bibfnamefont{M.}~\bibnamefont{Fabrizio}},
  \bibinfo{journal}{Phys. Rev. B} \textbf{\bibinfo{volume}{76}},
  \bibinfo{pages}{165110} (\bibinfo{year}{2007}).

\bibitem[{\citenamefont{B\"unemann et~al.}(1998)\citenamefont{B\"unemann,
  Weber, and Gebhard}}]{Gebhard}
\bibinfo{author}{\bibfnamefont{J.}~\bibnamefont{B\"unemann}},
  \bibinfo{author}{\bibfnamefont{W.}~\bibnamefont{Weber}}, \bibnamefont{and}
  \bibinfo{author}{\bibfnamefont{F.}~\bibnamefont{Gebhard}},
  \bibinfo{journal}{Phys. Rev. B} \textbf{\bibinfo{volume}{57}},
  \bibinfo{pages}{6896} (\bibinfo{year}{1998}).

\bibitem[{\citenamefont{Schir\`o and Fabrizio}(2010)}]{nostro_unpublished}
\bibinfo{author}{\bibfnamefont{M.}~\bibnamefont{Schir\`o}} \bibnamefont{and}
  \bibinfo{author}{\bibfnamefont{M.}~\bibnamefont{Fabrizio}},
  \bibinfo{journal}{in preparation}  (\bibinfo{year}{2010}).

\bibitem[{\citenamefont{Huber and R\"uegg}(2009)}]{Z2-1}
\bibinfo{author}{\bibfnamefont{S.~D.} \bibnamefont{Huber}} \bibnamefont{and}
  \bibinfo{author}{\bibfnamefont{A.}~\bibnamefont{R\"uegg}},
  \bibinfo{journal}{Phys. Rev. Lett.} \textbf{\bibinfo{volume}{102}},
  \bibinfo{pages}{065301} (\bibinfo{year}{2009}).

\bibitem[{\citenamefont{R\"uegg et~al.}(2010)\citenamefont{R\"uegg, Huber, and
  Sigrist}}]{Z2-2}
\bibinfo{author}{\bibfnamefont{A.}~\bibnamefont{R\"uegg}},
  \bibinfo{author}{\bibfnamefont{S.~D.} \bibnamefont{Huber}}, \bibnamefont{and}
  \bibinfo{author}{\bibfnamefont{M.}~\bibnamefont{Sigrist}},
  \bibinfo{journal}{Phys. Rev. B} \textbf{\bibinfo{volume}{81}},
  \bibinfo{pages}{155118} (\bibinfo{year}{2010}).

\bibitem[{\citenamefont{Das et~al.}(2006)\citenamefont{Das, Sengupta, Sen, and
  Chakrabarti}}]{Sengupta}
\bibinfo{author}{\bibfnamefont{A.}~\bibnamefont{Das}},
  \bibinfo{author}{\bibfnamefont{K.}~\bibnamefont{Sengupta}},
  \bibinfo{author}{\bibfnamefont{D.}~\bibnamefont{Sen}}, \bibnamefont{and}
  \bibinfo{author}{\bibfnamefont{B.~K.} \bibnamefont{Chakrabarti}},
  \bibinfo{journal}{Phys. Rev. B} \textbf{\bibinfo{volume}{74}},
  \bibinfo{pages}{144423} (\bibinfo{year}{2006}).

\bibitem[{\citenamefont{Anders et~al.}(2007)\citenamefont{Anders, Bulla, and
  Vojta}}]{spin-boson}
\bibinfo{author}{\bibfnamefont{F.~B.} \bibnamefont{Anders}},
  \bibinfo{author}{\bibfnamefont{R.}~\bibnamefont{Bulla}}, \bibnamefont{and}
  \bibinfo{author}{\bibfnamefont{M.}~\bibnamefont{Vojta}},
  \bibinfo{journal}{Phys. Rev. Lett.} \textbf{\bibinfo{volume}{98}},
  \bibinfo{pages}{210402} (\bibinfo{year}{2007}).

\end{thebibliography}

\end{document}